\documentclass[a4paper,10 pt]{article}
\usepackage[dvips]{graphicx}
\usepackage{amsmath}
\usepackage{amsfonts}
\usepackage{amssymb}
\usepackage{longtable}   
\allowdisplaybreaks[4]     
\usepackage[american,english]{babel}
\usepackage{textcomp}
\usepackage{fullpage}
\newcommand\beq{\begin{equation}}
\newcommand\eeq{\end{equation}}
\newcommand\beqs{\begin{equation*}}
\newcommand\eeqs{\end{equation*}}
\newcommand\beqa{\begin{eqnarray}}
\newcommand\eeqa{\end{eqnarray}}
\newcommand\beqas{\begin{eqnarray*}}
\newcommand\eeqas{\end{eqnarray*}}
\begin{document}
\title{\textbf{Fourth Order Gravity, Scalar-Tensor-Vector Gravity,
and Galaxy Rotation Curves}}
\author{Priti Mishra and 
Tejinder P. Singh\\ \\
Department of Astronomy and Astrophysics, \\
Tata Institute of Fundamental Research, \\
Homi Bhabha Road,  Mumbai -  400005, Maharashtra, India}

\date { }
\maketitle 
\begin{abstract}
\noindent
The Lambda-CDM model is the best fit to cosmological data, and to the observed galactic rotation curves. However, in the absence 
of a direct detection of dark matter one should explore  theories such as MOND, and perhaps also modified gravity theories like  
fourth order gravity and Scalar-Tensor-Vector Gravity [STVG] as possible explanations for the non-Keplerian behaviour of galaxy 
rotation curves. STVG has a modified law for gravitational acceleration which attempts to fit data by fixing two free parameters.
We show that, remarkably, the biharmonic equation which we get in the weak field limit of the field equations in a fourth order 
gravity theory implies a modification of Newtonian acceleration which is precisely of the same repulsive Yukawa form as in the 
STVG theory, and the corrections could in principle be large enough to try and explain the observed rotation curves.  We also 
explain how our model provides a first principles understanding of MOND. We also show that STVG and fourth order gravity predict 
an acceleration parameter $a_0$ whose value is of the same order as in MOND.
\end{abstract}

 
\section{Introduction}
 In spiral galaxies the observed rotation curve profiles are strongly  inconsistent  with those
predicted in Newtonian gravity and Galilean acceleration from the
distribution of the "luminous" matter we detect.
Dark matter is regarded as the most plausible explanation for the observed galaxy rotation curves. It is also strongly 
indicated via its pivotal role in the formation of large scale structures in the Universe, providing us with the standard 
$\Lambda$-CDM model.

However, until direct detection of one or more dark matter candidates takes place in the laboratory or through astronomical 
observations, it is perhaps useful to consider modified theories of gravity as alternate explanations for the observed galaxy 
rotation curves, and explore to what extent and to what accuracy they fit data. Such an analysis should be looked at in the same 
spirit in which modifications to general relativity such as $f(R)$ gravity are being considered as alternatives to dark energy 
and the cosmological constant, when it comes to explaining the acceleration of the Universe. It would be all the more appealing 
if a single modified gravity could account for rotation curves, cosmic acceleration and also satisfactorily explain structure 
formation. A preliminary attempt in this direction was made by us in \cite{PMTP2012}
using a specific fourth order modified gravity. In the present article we provide details of the analysis of this fourth order 
theory relevant to galactic rotation curves. 

Fourth order gravity theories have indeed been used before to explain rotation curves, but to the best of our knowledge, the field
equations proposed by us below [and which lead to a biharmonic equation in the weak field limit] have not been considered before. 
Earlier works on fourth order gravity in this context include those of Stabile and Scelza \cite{Stabile}, and Mannheim and 
Kazanas \cite{Mannheim}. The work \cite{Stabile} is based on a generalization of general relativity where the Einsten-Hilbert 
Lagrangian is replaced by a generic function of the Ricci scalar and the Ricci and Riemann tensors.
The work in \cite{Mannheim} is based on conformal Weyl gravity. Rotation curves in $R^n$ gravity have been considered in 
\cite{Martins}. Of course the fact that non-minimally coupled theories of gravity show a modification of Poisson's equation is 
well-known since at least thirty years (see for example the work of \cite{Stelle}).
 
There are various theoretical reasons [independent of the need to explain rotation curves or cosmic acceleration] for
 considering alternatives to general relativity [GR], including higher derivative theories of gravity. GR admits gravitational
 singularities, which are possibly avoided in a quantum theory of gravity.  Effective quantum corrections arising from quantum
 gravity theories can generically be interpreted as higher derivative corrections to GR. Furthermore, higher derivative theories
 are generally better behaved in the ultra-violet regime and allow for an improved possibility of constructing a singularity free
 gravity theory \cite{Biswas}. Higher order corrections to GR also arise from considerations of unification of interactions \cite{Will}.
 
 The form of field equations considered by us below is motivated by [though, is independent of]  a study of averaging of microscopic
 Einstein equations over a gravitationally polarized region, wherein fourth order effective corrections to Einstein equations appear
 owing to the existence of an underlying quadrupole moment in the mass distribution \cite{Szekeres}, \cite{Zalaletdinov}. Presently 
 however, we regard our fourth order equations as arising not from averaging over moments, but as an effective classical relic of an
 underlying quantum theory of gravity.  
 
 We first present the field equations and their solution in the weak field limit. Next, we take recourse to an earlier tentative solution
 for the rotation curves problem suggested by Moffat and collaborators in their Scalar-Tensor-Vector gravity, and show that this solution
 holds for our model too. Lastly, we compare our work with the solution for rotation curves proposed in MOND. 

\section{Fourth order gravity and the biharmonic equation for the potential}
A fourth order modified gravity has been postulated in ~\cite{PMTP2012} as a common explanation for the observed cosmic acceleration
and for galaxy rotation curves. Here we specialize the theory to the case of galactic dynamics and give the details of the analysis
pertinent to the structure of the rotation curves. The modified gravity is assumed to be described by the following effective field 
equations
\begin{equation}
R^{\mu\nu} - \frac{1}{2} g^{\mu\nu} R = \frac{8\pi G}{c^4} T^{\mu\nu} + 
k^{-2} R^{\mu\alpha\nu\beta}_{\ \ \ \ \ ;\alpha\beta}
\label{modee}
\end{equation}
where $k$ is a positive constant with dimensions of inverse length.

We shall be interested in the Newtonian weak-field limit of the above equations, so that we choose the metric to be
\begin{equation}
ds^2 =  \left( 1 + \frac{2\Phi}{c^2}\right) c^{2} dt^2 - dx^2 - dy^2 - dz^2\ 
\label{weakfd}
\end{equation}
and the only non-zero component of the energy-momentum tensor is $T^{0}_{0}=c^2 \mu(r)$ where $\mu(r)$ is the matter density distribution.

In this limit the modified field equations (\ref{modee}) reduce to the 
following fourth order biharmonic equation, which is a modification of the Poisson equation
\beq 
\nabla^4\Phi-k^2\nabla^2\Phi=-4\pi Gk^2\mu(r)
\label{original biharmonic eqn}
\eeq 
It will be shown that this
fourth order biharmonic modification of the Poisson equation explains the observed galaxy rotation curves without dark matter.

We are interested here in the modification
 of the radial dependence of the potential. Using separation of variables, the radial 
part of this equation is given by:
\begin{equation}
\label{bhradial}
\phi'''' + \frac{4}{r}\phi'''-k^{2}\phi''-\frac{2}{r}k^{2}\phi'=-4\pi Gk^{2}\mu(r)
\end{equation}
 where a prime denotes a derivative with respect to $r$.

We find the series solution of this equation using the standard Frobenius method around 
the regular singular point $r=0$.
\smallskip

\noindent \textbf{Case 1}: The vacuum solution $\mu=0$  [homogeneous equation]: 
In this case we get the following solution for the  acceleration $a=-\nabla\phi$
\beq 
a(r)=-\bigg(C_0+\frac{C_1}{k}\bigg)\frac{e^{kr}}{2kr}+\bigg(C_0-\frac{C_1}{k}\bigg)\frac{e^{-kr}}{2kr}+
\bigg(C_0+\frac{C_1}{k}\bigg)\frac{e^{kr}}{2k^2r^2}+\bigg(C_0-\frac{C_1}{k}\bigg)\frac{e^{-kr}}
{2k^2r^2}-\frac{C_2 }{r^2}
\label{19eqn}
\eeq 
Since we have assumed $k>0$, this implies that in Eqn. (\ref{19eqn}), terms proportional to
 $e^{kr}\rightarrow\infty$ as $r\rightarrow\infty$ which is unphysical. So we set the 
coefficient of $e^{kr}$ to zero. Thus
\beq 
\bigg(C_0+\frac{C_1}{k}\bigg)=0\Rightarrow C_1=-kC_0.
\eeq
Hence
\beq
a=C_0\frac{e^{-kr}}{kr}+C_0\frac{e^{-kr}}{k^2r^2}-\frac{C_2 }{r^2}
\eeq
\begin{center}
\fbox{\parbox{10cm}
{\beq 
a=-\frac{C_2 }{r^2}+C_0\frac{e^{-kr}}{k^2r^2}(1+kr)
\label{20eqn}
\eeq }}
\end{center}
Eqn. (\ref{20eqn}) is the solution for acceleration for the vacuum [homogeneous] case. 
The 
constants $C_0$ and $C_2$ can be related by the following reasoning: For $kr\ll 1$ 
we assume Newton's law of gravitation to hold, so that $C_2 = GM + C_{0}/k^{2}\equiv 
G_{\infty}M$ where $G_{\infty}=G [ 1 + C_{0}/k^{2} M G]$. For $kr\gg 1$ the exponential 
term can be ignored, and $G_{\infty}$ represents the effective gravitational constant 
at large distances. We note that overall there are two constants in the solution, $k$ and $C_0$, the former from the field equations,
and the latter as a constant of integration.

\smallskip

\noindent\textbf{Case 2}: $\mu(r)\neq 0 $ [Inhomogeneous Solution]:

\noindent Since we know the solution for the homogeneous case ($\mu=0$), we can construct 
the solutions for the inhomogeneous case using the homogeneous solutions. As is well known, if $y=au+bv$ is a solution of 
\beq 
y''+P(x)y'+Q(x)y=0
\label{21eqn}
\eeq 
where $a$ and $b$ are constants, we can find the solution of 
\beq 
y''+P(x)y'+Q(x)y=R(x)
\label{22eqn}
\eeq in the form 
\beq 
y=A(x)u+B(x)v
\label{23eqn}
\eeq 
 where \beq 
 A(x)=-\int \frac{vR}{W}dx,\newline
 B(x)=\int\frac{uR}{W}dx
 \label{24eqn}
 \eeq 
where $W$ is the Wronskian
\beq
W=uv'-u'v.
\eeq
Here,
\beq 
A(r)=4\pi Gk\int r\sinh(kr)\mu(r) dr+a_1,
B(r)=-4\pi Gk^2\int r\cosh(kr)\mu dr+a_2
\label{25eqn}
\eeq 
Hence inside a medium with a density profile $\mu(r)$ the acceleration is given by 
\beq
a=-\phi'=-\frac{1}{r^2}\int r^2\bigg[\frac{4\pi Gk\cosh(kr)}{r}\int r\sinh(kr)\mu dr-
$$ $$\frac{4\pi Gk\sinh(kr)}{r}\int r\cosh(kr)\mu dr+a_1\frac{\cosh(kr)}{r}+a_2\frac{\sinh(kr)}{kr}\bigg] dr-\frac{a_3 }{r^2}
\label{27eqn}
\eeq 
 The last two terms in the bracket in the above equation are same as in the homogeneous case. These two will be reduced to 
${a_1e^{-kr}}/{k^2r^2}(1+kr)$ for the same reasons as discussed before Eqn. (\ref{20eqn}).
Hence Eqn. (\ref{27eqn}) becomes
\begin{center}
\fbox{\parbox{15cm}
{\beq
a=-\phi'=-\frac{1}{r^2}\int r^2\bigg[\frac{4\pi kG\cosh(kr)}{r}\int r\sinh(kr)\mu dr-
$$ $$\frac{4\pi Gk\sinh(kr)}{r}\int r\cosh(kr)\mu dr\bigg] dr+\frac{a_1e^{-kr}}{k^2r^2}(1+kr)-\frac{a_3 }{r^2}
\label{28eqn}
\eeq }}
\end{center} 
Eqn. (\ref{28eqn}) gives the acceleration for any given $\mu(r)$. One can easily read off special cases from here :

\textbf{Case I} : $\mu=0$ :
\beq 
a=-\phi'=\frac{a_1e^{-kr}}{k^2r^2}(1+kr)-\frac{a_3 }{r^2}
\eeq
which is same as equation (\ref{20eqn}), as expected. Note that $a_3$ is related to $a_1$ and $k$ as mentioned below (\ref{20eqn}).

\textbf{Case II}: $\mu=$constant=$\mu_0$ :
\beqs 
a= -\phi'=-\frac{4\pi G\mu_0}{r^2}\int \bigg[ \frac{r^2k\cosh(kr)}{r}
\left\{\int r\sinh(kr)dr\right\}\bigg]dr+
\eeqs
\beq
\frac{4\pi G\mu_0}{r}\int \bigg[\frac{r^2k\sinh(kr)}{r}\left\{\int r\cosh(kr)dr\right\}\bigg]dr
+\frac{a_1e^{-kr}}{k^2r^2}(1+kr)-\frac{a_3 }{r^2}
\eeq
or 
\begin{center}
\fbox{\parbox{10cm}
{\beq 
a=-\frac{4\pi G\mu_0}{3}r+\frac{a_1e^{-kr}}{k^2r^2}(1+kr)-\frac{a_3 }{r^2}
\label{29eqn}
\eeq  }}
\end{center}
Eqn. (\ref{29eqn}) gives acceleration for a medium of constant density. Once again, $a_3$ is related to $a_1$ and $k$, and the same
holds for the general solution (\ref{28eqn}). So in all the cases there are two free constants, which are to be fixed by comparison
with observations of galactic rotation curves.

We now need to know the solution of the integrals in the general expression (\ref{29eqn}) for the acceleration, once a density profile
$\mu(r)$ is given. This is a difficult task. However, we can put to use the results discovered earlier in a modified 
gravity theory known as Scalar-Tensor-Vector Gravity (STVG) developed by Moffat and 
collaborators ~\cite{Moffat1,BrownsteinandMoffat,BrownsteinThesis}. It turns out that a solution for the velocity profile, found in
STVG for possibly fitting galactic rotation curves, is also a solution in our case, for realistic density profiles. 

Below we briefly recall how the acceleration and velocity profile in STVG is arrived at, and in the subsequent section we use the 
results of STVG for our fourth order gravity.

\section{Galaxy Rotation Curves in the STVG Theory}

Here we briefly recall the form of the acceleration law, as derived in STVG theory. Details can be found 
in ~\cite{Moffat1,BrownsteinandMoffat,BrownsteinThesis}.  For a related recent discussion 
see ~\cite{Rahvar}.

The STVG theory includes, apart from standard gravity, a massive vector field $\phi^{\mu}$ with a coupling $\omega$ to gravity
and mass $\mu$. In its most general form, STVG assumes the coupling
$\omega$, the mass $\mu$ and the gravitational constant $G$ to be space-time dependent scalar fields
$\omega(x)$, $\mu(x)$ and $G(x)$ respectively. In STVG, the action is given by
\begin{equation}
S=S_{\rm Grav}+S_\phi+S_S+S_M,
\end{equation}
where
\begin{equation}
S_{\rm Grav}=\frac{1}{16\pi}\int d^4x\sqrt{-g}\biggl[
\frac{1}{G}(R+2\Lambda)\biggr],
\end{equation}
\begin{equation}
S_\phi=-\int
d^4x\sqrt{-g}\biggl[\omega\biggl(\frac{1}{4}B^{\mu\nu}B_{\mu\nu} +
V(\phi)\biggr)\biggr],
\end{equation}
and
\begin{equation}
\label{Saction} S_S=\int
d^4x\sqrt{-g}\biggl[\frac{1}{G^3}\biggl(\frac{1}{2}g^{\mu\nu}\nabla_\mu
G\nabla_\nu G-V(G)\biggr)
$$ $$
+\frac{1}{G}\biggl(\frac{1}{2}g^{\mu\nu}\nabla_\mu\omega
\nabla_\nu\omega-V(\omega)\biggr)
+\frac{1}{\mu^2G}\biggl(\frac{1}{2}g^{\mu\nu}\nabla_\mu\mu\nabla_\nu\mu
-V(\mu)\biggr)\biggr].
\end{equation}
Here  $V(\phi)$ is the
potential for the massive vector field $\phi^\mu$ having mass $\mu$ and coupling $\omega$; $V(G), V(\omega)$
and $V(\mu)$ are the potentials associated with the three
scalar fields $G(x),\omega(x)$ and $\mu(x)$, respectively, and
\begin{equation}
B_{\mu\nu}=\partial_\mu\phi_\nu-\partial_\nu\phi_\mu.
\end{equation}

Next, considerable simplification is imposed on the model. By assuming that $\Lambda=0$, $V(\phi)=0$, $\omega$ and $\mu$ to be
constants, and by considering the 
motion of a test particle coupled to gravity and to the vector field, it was shown that the law of acceleration for the field 
outside a spherical mass $M$ is given by
\begin{equation}
a(r)=-\frac{G_{\infty}M}{r^2}+K\frac{\exp(-r/r_0)}{r^2}\biggl(1+\frac{r}{r_0}\biggr),
\label{moffatlaw}
\end{equation}
where $G_{\infty}$ is defined to be the effective gravitational
constant at infinity
\begin{equation}
\label{renormG}
G_{\infty}=G\biggl(1+\sqrt{\frac{M_0}{M}}\biggr)
\end{equation}
and $r_0=1/\mu$.
Here, $M_0$ denotes a parameter that vanishes when $\omega=0$ and
$G$ is Newton's gravitational constant. The constant
$K$ is assumed to equal
\begin{equation}
\label{Kequation} K=G\sqrt{MM_0}.
\end{equation}
The choice of $K$, which determines the strength of the coupling
of $B_{\mu\nu}$ to matter and the magnitude of the Yukawa force
modification of weak Newtonian gravity, is based on phenomenology
and $M_0$ is a free parameter of the theory. This particular choice of expression for $K$ ensures that
for $r\ll r_0$ the acceleration law reduces to $a(r) = - GM/r^2$, as may be verified by expanding the exponential in the
acceleration law (\ref{moffatlaw}).

By using (\ref{renormG}), one can rewrite the acceleration in the
Yukawa form
\begin{equation}
\label{accelerationlaw} a(r)=-\frac{G
M}{r^2}\biggl\{1+\sqrt{\frac{M_0}{M}}\biggl[1-\exp(-r/r_0)
\biggl(1+\frac{r}{r_0}\biggr)\biggr]\biggr\}.
\end{equation}
One can generalize this to the case of interior of a mass distribution by
replacing the factor $GM/r^2$ in (\ref{accelerationlaw}) by
$G{ M}(r)/r^2$:
\begin{equation}
\label{accelerationlawinterior} a(r)=-\frac{G
{ M}(r)}{r^2}\biggl\{1+\sqrt{\frac{M_0}{M}}\biggl[1-\exp(-r/r_0)
\biggl(1+\frac{r}{r_0}\biggr)\biggr]\biggr\}.
\end{equation}
 The rotational velocity of a star $v_c$ is
obtained from $v_c^2(r)/r=a(r)$ and is given by
\begin{equation}
v_c=\sqrt{\frac{G{
M}(r)}{r}}\biggl\{1+\sqrt{\frac{M_0}{M}}\biggl[1-\exp(-r/r_0)
\biggl(1+\frac{r}{r_0}\biggr)\biggr]\biggr\}^{1/2}.
\end{equation}

Moffat and Brownstein \cite{BrownsteinandMoffat} used the data of K and B- band luminosity,
velocity, distance and redshift  of these galaxies from the works of 
Avilla-Reese et al. \cite{Avila}, Balin et al. \cite{Balin}, Begeman et al. \cite{Begeman} and
Bekenstein \cite{Bekenstein}. Moffat and Brownstein obtained reasonably useful parametric fits to the rotation curves of 
101 galaxies from this data-set - these rotation curves are available in Fig. 2 of their paper 
\cite{BrownsteinandMoffat} (here in our paper we have used a subset of this very dataset). To our understanding, their 
analysis and its results, which are also arrived at by us here, come close to achieving
the Universal Rotation Curve proposed by Salucci et al. \cite{Salucci}.
A good fit to a large number of galaxies has been
achieved with the parameters:
\begin{equation}
M_0=9.60\times 10^{11}\,M_{\odot},\quad r_0=13.92\,{\rm
kpc}=4.30\times 10^{22}\,{\rm cm}.
\end{equation}
In the fitting of the galaxy rotation curves for both LSB and HSB
galaxies, using photometric data to determine the mass
distribution ${M}(r)$,  only the
mass-to-light ratio $\langle M/L\rangle$ is employed, once the
values of $M_0$ and $r_0$ are fixed universally for all LSB and
HSB galaxies. Dwarf galaxies are also fitted with the
parameters
\begin{equation}
M_0=2.40\times 10^{11}\,M_{\odot},\quad r_0=6.96\,{\rm
kpc}=2.15\times 10^{22}\,{\rm cm}.
\end{equation}

One can criticize the use of different values of $M_0$ and $r_0$ for different classes of galaxies, which takes away
any possibility of them being universal numbers; and instead being chosen at will to fit observations. On the one 
hand such criticism might appear well-justified; on the other hand a closer examination reveals the possibility of 
profound underlying physics whose origins are yet to be understood. The ratio $GM_0/r_0^2$ for HSB and LSB galaxies 
is the same as for dwarf galaxies, and this ratio is of the order of the observed cosmic acceleration  $cH_{0}^{-1}$ 
(and the MOND acceleration). 
The constancy of this ratio has been observed across a family of large-scale-structures (see the remarkable Fig. 7 
in \cite{Moffat:2009}) and formed the basis of our proposal \cite{PMTP2012} that cosmic acceleration and galactic
rotation curves can be explained by the same physical origin.   Also, these values of the
Yukawa parameters do not violate constraints coming from solar system and laboratory data [see \cite{Adelberger} 
and Fig. 8 of \cite{Moffat:2009}].

We will now compare the modified acceleration law of the form Eqn. (\ref{accelerationlawinterior}) with the law
obtained from the fourth order gravity, and show that the above interior solution works for the fourth order case as well.

\section{Comparison with the solution in Fourth Order Gravity} 

A comparison and realization of the similarity between our fourth order gravity solution and the STVG solution provides
 a useful hint that our solution is a useful candidate for understanding rotation curves without possibly invoking dark matter.

As noted above, in STVG the acceleration outside of a body is given by 
\beq
a(r)=-\frac{G_\infty M}{r^2}+\frac{K}{r^2} \exp(-\frac{r}{r_0})\bigg(1+\frac{r}{r_0}\bigg)
\label{30eqn}
\eeq
This form matches exactly with that derived in (\ref{20eqn}) for the biharmonic equation.  
Comparing Eqn. (\ref{20eqn}) with Eqn. (\ref{30eqn}),
\beqa
C_2&=&G_\infty M=G\left(1+\sqrt\frac{M_0}{M}\right) M,\\
\frac{C_0}{k^2}&=&K=G\sqrt{MM_0}\\
\& \hspace{2cm} k&=&\frac{1}{r_0}.
\eeqa

Next, in STVG theory, the acceleration inside a spherical mass distribution takes the form 

\beq 
a(r)=-\frac{GM(r)}{r^2}\bigg\{1+\sqrt{\frac{M_0}{M}}\bigg[1-\exp(-r/r_0)\bigg(1+\frac{r}{r_0}\bigg)\bigg] \bigg\}
\label{ouracc}
\eeq 
This is the form used to fit galaxy rotation curves. We write the above equation in the following form:
\beq 
a(r)=\frac{M(r)}{M}A(r)
\label{amog}
\eeq 
where 
\beq 
A(r)=-\frac{GM}{r^2}\bigg\{1+\sqrt{\frac{M_0}{M}}\bigg[1-\exp(-r/r_0)\bigg(1+\frac{r}{r_0}\bigg)\bigg] \bigg\}
\eeq 

Now one would like to check whether the biharmonic equation also gives the same form of acceleration 
inside a spherical mass distribution. In general this would be difficult to check; below we describe 
a way which suffices for our purpose. 

The solution of the biharmonic equation for any given mass distribution $\mu(r)$ is given by Eqn. (\ref{28eqn}). Rewriting 
(\ref{28eqn}),
\beq
a(r)=-\frac{1}{r^2}\int \bigg[4\pi kGr\cosh(kr)\int r\sinh(kr)\mu dr-$$ $$4\pi Gkr\sinh(kr)\int r\cosh(kr)\mu dr\bigg] dr+A(r)
\eeq 
with \beq 
a_1=k^2G\sqrt{MM_0},\hspace{0.1cm} a_3=G\sqrt{MM_0}+GM
\eeq 
 we see 
that there is double integral over $\mu(r)$ in the above equation. We construct a function of the acceleration and its 
derivatives which is independent of integrals. Such a function is 
\beq 
\chi(r)=\beta''(r)-k^2\beta(r)
\eeq 
where 
\beq 
\beta=\frac{1}{r}\frac{d}{dr}[r^2(a(r)-A(r))]
\eeq 
From the solution of the biharmonic equation it can be shown that (as expected)
\beq 
\chi=4\pi Gk^2r\mu(r)
\eeq 
We also find $\chi(r)$ from STVG theory also using equation(\ref{amog}). Now 
\beq 
a_{STVG}(r)=a_{bh}(r),
\eeq 
\beq 
\Rightarrow \chi_{STVG}(r)=\chi_{bh}(r)
\eeq 
where  \textquotedblleft bh\textquotedblright stands for biharmonic equation. Equating $\chi_{STVG}$ 
and $\chi_{bh}$, we get an equation of the 
form
\beq
P(r)M(r)+Q(r)M'(r)+R(r)M''(r)+S(r)M'''(r)+T(r)=0.
\label{2ndordermu}
\eeq
where
\beqa
P(r)&=&-\frac{e^{-r/r_0} G \sqrt{\frac{M_0}{M}}}{r_0^4},\\
Q(r)&=&G\bigg[-\frac{2}{r^3}+\sqrt{\frac{M_0}{M}}\bigg(-\frac{2}{r^3}+\frac{2e^{-r/r_0}}{r^3}+\frac{3e^{-r/r_0}}
{r_0^3}+\frac{e^{-r/r_0}}{rr_0^2}+\frac{2e^{-r/r_0}}{r^2r_0}\bigg)-\frac{4\pi}{rr_0^2}\bigg],\nonumber\\
R(r)&=&G\bigg(\frac{2}{r^2}+\frac{2\sqrt{\frac{M_0}{M}}}{r^2}-\frac{2e^{-r/r_0}\sqrt{\frac{M_0}{M}}}{r^2}-\frac{3e^{-r/r_0}
\sqrt{\frac{M_0}{M}}}{r_0^2}-\frac{2e^{-r/r_0}\sqrt{\frac{M_0}{M}}}{rr_0}\bigg),\\
S(r)&=& G\bigg(-\frac{1}{r}-\frac{\sqrt{\frac{M_0}{M}}}{r}+\frac{e^{-r/r_0}\sqrt{\frac{M_0}{M}}}{r}+
\frac{e^{-r/r_0}\sqrt{\frac{M_0}{M}}}{r_0}
\bigg)\\
\text{and }T(r)&=&-\frac{e^{-r/r_0} GM \sqrt{\frac{M_0}{M}}}{r_0^4}.
\eeqa
\beq
\because  M(r)=\frac{1}{r^2}\int \mu r^2dr,
\eeq 
Eqn. (\ref{2ndordermu}) contains one integral over $\mu(r)$. Dividing equation(\ref{2ndordermu}) 
by $P(r)$ and then differentiating, we get a
differential equation of the form,
\beq
U(r)\mu(r)+V(r)\mu'(r)+W(r)\mu''(r)+X(r)\mu'''(r)=0
\label{3rdordermu}
\eeq
where
\beqa
U(r)&=&\bigg(4 e^{r/r_0} M \sqrt{\frac{M_0}{M}} \pi  r_0 (r+r_0)+M_0 (r^2-6 r_0 r+5r_0^2)\bigg),\\
V(r)&=&r_0\bigg[2 e^{r/r_0} M \sqrt{\frac{M_0}{M}}r_0(2 \pi r+r_0)+M_0\bigg(-3 r^2+11 r_0 r+2 (-1+e^{r/r_0}) r_0^2\bigg)\bigg],\\ 
W(r)&=&r_0^2\bigg[e^{r/r_0} M \sqrt{\frac{M_0}{M}} r_0 (r+3 r_0)+M_0 \bigg(3 r^2+(-4+e^{r/r_0})rr_0
+3 (-1+e^{r/r_0}) r_0^2\bigg)\bigg], \nonumber \\
X(r)&=&-rr_0^3 \bigg[M_0 (r-e^{r/r_0}r_0+r_0)-e^{r/r_0} M \sqrt{\frac{M_0}{M}} r_0\bigg].
\eeqa
The solution of Eqn. 
(\ref{3rdordermu}) will tell us for what value of 
$\mu(r)$, STVG theory and the biharmonic equation give the  same acceleration inside a mass distribution. 
 But it is not necessary to solve (\ref{3rdordermu}). Instead we take the observed density 
profile $\mu(r)$ for a specific galaxy type and see whether it satisfies (\ref{3rdordermu}).
 
Following \cite{BrownsteinandMoffat} and known observational data, 
we assume the following density profile $\mu(r)$ 
\beq 
\mu(r)=\frac{3}{4\pi r^{3}} \beta  M(r) \bigg[\frac{r_c}{r+r_c}\bigg]
\eeq 
where
\beq 
M(r)=4\pi \int_0^rdr'r'^2\mu(r')=M\bigg(\frac{r}{r+r_c}\bigg)^{3\beta}
\eeq 
and
\beq 
\beta = \left\{
\begin{array}{ll} 1 & \mbox{for HSB galaxies,} \\
2 & \mbox{for LSB \& Dwarf galaxies.}
\end{array} \right.
\eeq 

The reader is referred to Section 2 of  \cite{BrownsteinandMoffat} for a detailed discussion of the assumptions
 behind the choice of the mass distribution assumed above, which is a simplified parametric model which includes the
 contribution coming from the stellar as well as the gaseous HI component, and  the choice of an assumed $M/L$ ratio. 

We have taken the data set of ten LSB, HSB and dwarf galaxy masses from Table (3) in ~\cite{BrownsteinandMoffat}. 
Then we plot 
the left hand side of Eqn. (\ref{3rdordermu}) 
\beq
\label{check}
STVG-BH\equiv \frac{r_0}{M_0^2}(U(r)\mu(r)+V(r)\mu'(r)+W(r)\mu''(r)+X(r)\mu'''(r))
\eeq
after making it dimensionless using parameters $r_0$ and $M_0$. We have made the assumption that $k=1/r_0$. 
We have taken data set of $10$ LSB, HSB and dwarf galaxies and for each of them 
we have computed $STVG-BH$ vs. $r$ for three ranges of $r$- ($0.0001-1$) kpc, ($1-100$) kpc and ($100-500$) kpc. 
We find that for the range $1-100$ kpc $STVG-BH$ is 
very small, of the order of $10^{-4}$ or $10^{-3}$. And for other ranges $STVG-BH$ takes high values. Since the 
range of length scales of interest in a galaxy lie between 
$1-100$ kpc and $STVG-BH$ is very small 
in this range, it can be said that the solution coming from STVG theory matches with the solution of the 
biharmonic equation, inside the medium,  for the observed density profile. In this sense, the modified 
acceleration law resulting from the biharmonic equation is also a candidate for explaining the observed rotation curves, 
from the dataset 
used by Moffat and Brownstein if $k$ and $M_0$ are assumed to take the values needed to explain observations. In future work we 
plan to explicitly investigate how close the predictions of the biharmonic equation 
can come to the Universal Rotation Curve proposed by Salucci et al. \cite{Salucci}.

The modified gravity law (\ref{ouracc}) predicted by our fourth order gravity and the biharmonic equation is not ruled out by
 laboratory and solar-system tests of a Yukawa type departure from the inverse square law. This maybe seen by writing 
Eqn.(\ref{ouracc}) in the commonly used Yukawa form
\beq
a(r) = - \frac{G_{Y} M(r)}{r^2} \left\{1 + \alpha_{Y} \left(1 +\frac{r}{\lambda} \right) 
e^{-r/\lambda}\right\}
\eeq
where
\beq
\alpha = \sqrt{M_0/M}, \qquad \alpha_Y = - \frac{\alpha}{1+\alpha}, 
\qquad G_Y = \frac{G}{1+\alpha_Y}, \qquad \lambda = r_0
\eeq
In our work the typical values for $\lambda$ and $|\alpha_Y|$ are $\lambda\approx 10^{20}$ m and 
$|\alpha_Y|$ is order unity. Various laboratory and astronomical tests have been carried out to date to constrain the values
 of these two parameters, which respectively measure the critical length scale over which the departure from inverse square 
law becomes significant, and the strength of the departure. These tests are discussed in detail by Talmadge et al. \cite{Talmadge},
 by Fischbach and Talmadge 
(Fig. 2.13) \cite{Fischbach}, by Adelberger et al. (Fig. 4) \cite{Adelberger} and by Moffat and Toth (Fig. 8) \cite{Moffat:2009}. 
These tests include analysis of planetary orbits and planetary precession, laboratory measurements of the gravitational constant, 
geophysical tests, orbital motion of the LAGEOS satellite and the moon, and Lunar Laser Ranging (anomalous perigee precession). As 
is evident from these discussions, and especially from Fig. 8 of \cite{Moffat:2009} these tests do not at present reach out to the 
range necessary to rule out the parameter values required in our fourth order theory to explain galaxy rotation curves.

\section{Comparison of our work with MOND}
Modified Newtonian Dynamics (MOND) ~\cite{Migrom1}, ~\cite{Migrom2} and ~\cite{Migrom3}
  is a hypothesis that proposes a modification of Newton's law of gravity to explain the 
observed non-Newtonian galaxy 
rotation curves. In MOND, the gravitational force acting on a test particle of mass $m$ is assumed to be given by
the relation
\beq 
F=m a\mu\bigg(\frac{a}{a_0}\bigg)
\eeq 
where $a$ is acceleration in Newtonian mechanics and $a_0$ is a new fundamental constant of nature having the value 
$2\times 10^{-10}$ ms$^{-2}$. 
For very small accelerations [large distances] it is assumed that 
\beq 
\mu\bigg(\frac{a}{a_0}\bigg)=\frac{a}{a_0}
\eeq 
whereas $\mu$ approaches the value one for accelerations encountered in the solar neighborhood. For a detailed discussion of MOND,
 including possible functional forms for $\mu$ see for instance ~\cite{bekenstein}.

Hence for large distances we will have the relation 
\beqa 
\frac{GM}{r^2}=\frac{a^2}{a_0}\\
\Rightarrow a=\frac{\sqrt{GMa_0}}{r}.\label{MONDacceleration}
\eeqa  
The virtue thus is that $a$ falls as $1/r$ rather than $1/r^2$. Thus equating $a$ to the centripetal acceleration $v^2/r$ we get that
\beq 
\frac{v^2}{r}=a=\frac{\sqrt{GMa_0}}{r}\quad \quad \\
\Rightarrow\quad  v=(GMa_0)^{1/4}.
\eeq 
With the numerical choice of $a_0$ made above one gets the desired  value of the velocity. 

As we have seen above, in MOND the law of motion under a gravitational force is modified but the law of gravity is not changed. 
However in our work and in STVG as well the effective law of gravitation is modified by a Yukawa potential while the law of motion
 is unchanged. Essentially we can say that for us 
\beq 
F=m a = \frac{G M}{r^2}\frac{1}{\mu}.
\eeq 

In our work and in STVG theory as well, the acceleration is of the form 
\beq 
a(r)=\frac{a_N }{\mu}
\eeq 
where \beq 
a_N =\frac{GM}{r^2}
\eeq 
and \beq 
\frac{1}{\mu}=\frac{M(r)}{M}A(r).
\eeq 
We will now see in what sense this form matches with MOND in the 
region of interest.
\begin{figure}[!htb]
\begin{center}
\includegraphics[width=\textwidth]{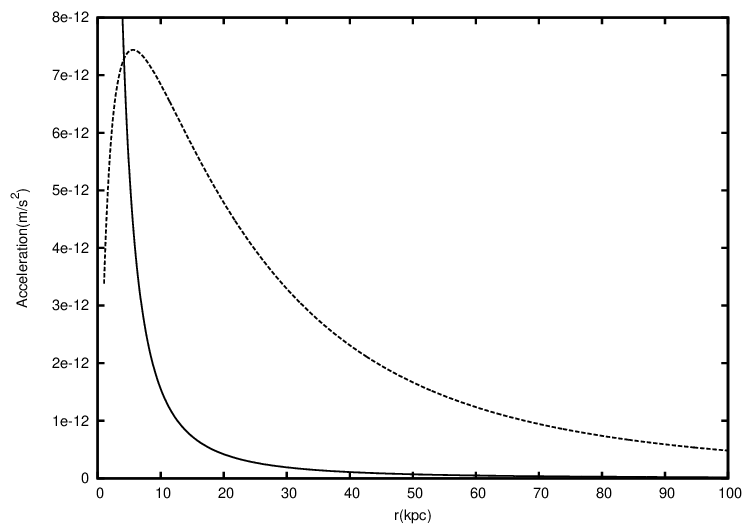}
\caption{Correction to the inverse square law as described by Eqn. (\ref{accelrationintwoparts}): the solid curve 
depicts the Newtonian fall-off given by the first term of (\ref{accelrationintwoparts}), whereas the dashed curve depicts the correction 
given by the second term of (\ref{accelrationintwoparts}).}
\label{1standsecondterminacceleration}
\end{center}
\end{figure}
\begin{figure}[!htb]
\begin{center}
\includegraphics[width=\textwidth]{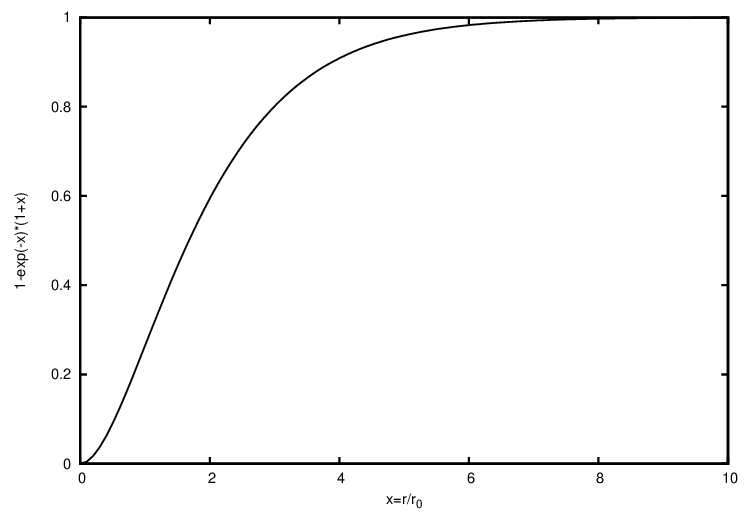}
\caption{Yukawa type term in acceleration: plot of the correction factor in Eqn. (\ref{STVGacc2})}
\label{2ndterm}
\end{center}
\end{figure}
\begin{figure}[!htb]
\begin{center}
\includegraphics[width=\textwidth]{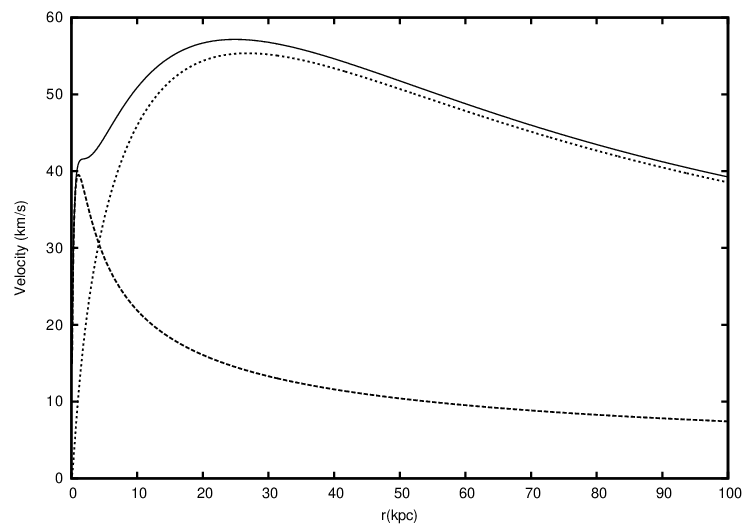}
\caption{Plot of the velocity profile corresponding to the acceleration given by Eqn. (\ref{accelrationintwoparts}). 
Dashed curve: velocity due to 1st term, dotted curve velocity due to 2nd term, thin curve: total velocity}
\label{total velocity}
\end{center}
\end{figure}

We have 
\beq
a(r)=-\frac{GM(r)}{r^2}\bigg\{1+\sqrt{\frac{M_0}{M}}\bigg[1-\exp(-r/r_0)\bigg(1+\frac{r}{r_0}\bigg)\bigg] \bigg\} \label{STVGacceleration}
\eeq
Let us write,
\beqa 
\frac{r}{r_0}&=&x\\
\Rightarrow a(r)&=&-\frac{GM(r)}{r_0^2x^2}\bigg[1+\sqrt{\frac{M_0}{M}}\bigg\{1-\exp(-x)\bigg(1+x\bigg)\bigg\} \bigg] \label{STVGacc2}
\eeqa 
We consider the region $r\gtrsim r_0$ and we define 
\beq
x=1+y,
\eeq
where $y$ is small and $y>0$.

Hence the term in square brackets in the expression of $a(r)$ can be written as,
\beqa 
&=&\bigg[1+\sqrt{\frac{M_0}{M}}\bigg\{1-\exp(-x)(1+x)\bigg\} \bigg] \\
&=& \bigg[1+\sqrt{\frac{M_0}{M}}\bigg\{1-\exp(-1-y)(2+y)\bigg\} \bigg]\\
&=&\bigg[1+\sqrt{\frac{M_0}{M}}\bigg\{1-\frac{2+y}{e}e^{-y}\bigg\}\\
\eeqa
\text{This essentially is $1/\mu$ in the MOND notation.}
(\text{Since $y$ is very small, $e^{-y}\approx1-y$}) and we can write this equation as\\
\beqa
&\cong& \bigg[1+\sqrt{\frac{M_0}{M}}\bigg\{1-\frac{(2+y)(1-y)}{e}\bigg\}\\
&\cong&\bigg[1+\sqrt{\frac{M_0}{M}}\bigg\{1-\frac{(2-y)}{e}\bigg\} \\ 
\eeqa
\text{Keeping terms upto only first order in y}\\
\beqa
&\cong&\bigg[1+\sqrt{\frac{M_0}{M}}\bigg\{1-\frac{2}{e}+\frac{y}{e}\bigg\}.
\eeqa 
$\Rightarrow$ for $r\gtrsim r_0$,
 \beq
a(r)\approx - \frac{GM(r)}{r^2}\bigg[1+\sqrt{\frac{M_0}{M}}\bigg\{1-\frac{2}{e}+\frac{y}{e}\bigg\}\bigg].
\eeq 

We write this acceleration as a sum of two parts, one falling as $1/r^2$ and another falling as $1/r$ :

\beq
a(r)\approx - \frac{GM(r)}{r^2}\bigg[1+\sqrt{\frac{M_0}{M}}\bigg\{1-\frac{3}{e}\bigg\}\bigg]- \frac{GM(r)}{r}\bigg[
\sqrt{\frac{M_0}{M}}\frac{1}{r_0 e}\bigg].
\label{accelrationintwoparts}
\eeq 

 Figure \ref{1standsecondterminacceleration} compares the first and second terms in the acceleration formula given above. As anticipated 
the second term dominates for large $r$. Also, for the acceleration to behave as $1/r$ in the region of interest, the second term 
inside the curly braces in Eqn. (\ref{STVGacceleration}) should rise linearly with $r$ - this is evident from Fig.\ref{2ndterm}. 
The region in which the second term dominates is where the galaxy rotation curve becomes non-Newtonian. This is clear from Fig. 
\ref{total velocity}. 
 At even larger distances, the second term is exponentially damped and the fall off is again
 Keplerian but with an effectively larger value of $G$.

In summary it is evident that in our work the observed rotation curve arises because the Yukawa type correction dominates over the
monopole term in the middle region, whereas the monopole dominates at either end.

It is significant that from our work we can give an estimate of the theoretical value of $a_0$ in MOND. 
A simplistic guess would be to construct a quantity 
with the dimension of acceleration from our fundamental quantities $k$ and $M_0$. This is nothing but 
\beqa
GM_0k^2=\frac{GM_0}{r_0^2}=&=&\frac{6.67\times 10^{-11}\times 9.60\times10^{11}\times 2\times 
10^{30}}{(13.92\times 10^3\times3.08\times 10^{16})^2}\\
&=&3\times 10^{-10} \ \rm{ms^{-2}}
\eeqa 
which is of the same order as $a_0$ in MOND. ( $a_0=2\times 10^{-10}$m/s$^2$). The relation is made more transparent by comparing 
the acceleration given by the second term of Eqn. (\ref{accelrationintwoparts}) with the acceleration in MOND in 
Eqn. (\ref{MONDacceleration}). This comparison yields the fundamental relation 
\beq 
a_0=\frac{GM_0}{r_0^2e^2}
\eeq 
 It shows that the parameter $a_0$ in MOND is of the same order as predicted from fourth order gravity.
 In principle then, the MOND effect can alternately be attributed to a modified fourth order gravity. The 
 difference from MOND is that in the modified gravity case the rotation curve again falls at very large distances,
 whereas in MOND it continues to remain flat. 
 
\section{Conclusions}
We found the general series solution to the biharmonic equation for the gravitational potential in the proposed fourth order gravity. 
We saw that the form of acceleration is very similar to that in STVG ~\cite{Moffat1} in vacuum. Then we investigated further if the 
form of acceleration matches inside the matter distribution also. What we find is that both the solutions may be successful candidates
 for the observed matter density profile of galaxies. As a consequence our fourth order gravity model is capable of explaining the 
observed rotation curves.

The fact that the modified acceleration law discussed in this paper also arises from apparently 
independent considerations as in STVG and MOND, is intriguing and suggests that further investigation would 
be of interest.

\bigskip

\noindent{\bf Acknowledgements} : One of the authors (TPS) is grateful to Roustam Zalaletdinov for 
extensive discussions and collaboration during the early part of this work, and for telling him 
about the work of Szekeres, and his own work, on gravitational quadrupole polarization. The 
solution  (\ref{20eqn}) of the biharmonic equation was found jointly with RZ, during the latter's visit 
to TIFR. 
Thanks are due to Aniket Agrawal, J. R. Brownstein, John Moffat and Viktor T. Toth for useful conversations, and also to
 Eric Adelberger, Marco Bruni, Patrick DasGupta, Eiichiro Komatsu and Cenalo Vaz.

\end{document}